\documentclass[12pt]{article}
\usepackage{graphicx}

\def\la{\mathrel{\mathpalette\fun <}}
\def\ga{\mathrel{\mathpalette\fun >}}
\def\fun#1#2{\lower3.6pt\vbox{\baselineskip0pt\lineskip.9pt
\ialign{$\mathsurround=0pt#1\hfil##\hfil$\crcr#2\crcr\sim\crcr}}}

\title{\bf Atomic levels in superstrong magnetic fields and
$D=2$ QED of massive electrons: screening}
\author{M.I. Vysotsky\\
~~ \\ {\em ITEP, Moscow}}

\begin{document}
\maketitle

\begin{abstract}

The photon polarization operator in superstrong magnetic fields
induces the dynamical photon ``mass'' which leads to screening of
Coulomb potential at small distances $z\ll 1/m$, $m$ is the mass 
of an electron. We demonstrate
that this behaviour is qualitatively different from the case of
$D=2$ QED, where the same formula for a polarization operator
leads to screening at large distances as well. Because of screening the
ground state energy of the hydrogen atom at the magnetic fields $B
\gg m^2/e^3$ has the finite value $E_0 = -me^4/2 \ln^2(1/e^6)$.

\end{abstract}

\section{Introduction}

The Larmour radius of the electron orbit $a_H = 1/\sqrt{eB}$ is much
smaller than Bohr atomic radius $a_B = 1/(me^2)$ for homogenius
magnetic fields $B\gg m^2 e^3$ (we are using Gauss system of
units, where $e^2 = \alpha = 1/137$; also in all formulas $\hbar =
c = 1$). It is natural to look for the atomic energy levels in such strong magnetic fields studing the influence of Coulomb potential on the
electrons occupying Landau levels \cite{Lan}.
 A strong magnetic field confines an electron in the
transverse direction while in the longitudinal direction an
electron is bound by the weak Coulomb field of a nucleus. The
cigar-shape wave function of an electron is formed with transverse
size which equals Larmour radius and longitudinal size which is by
$\ln(a_B^2/a_H^2) \equiv \ln(B/m^2 e^3)$ smaller than Bohr radius.
The ground state energy is larger than Rydberg constant by the
square of the same logarithm: $E_0 = -(me^4/2)\ln^2(B/m^2 e^3)$.
One can easily get this logarithmic factor from the fact that in
one-dimensional Coulomb potential energy diverges logarithmically
at small distances. The divergency is regularized  at the
longitudinal distances which equals $a_H$, where one-dimensional
motion converts to a three-dimensional one.
Atomic levels in such strong magnetic fields were found numerically 
in \cite{1} (see also \cite{2, 22}).

Our purpose is to understand the behaviour of the energy levels
with the growth of a magnetic field. The point is that at
superstrong magnetic fields $B\ga m^2/e^3$ the polarization
operator insertions into the photon propagator induce the dynamic
photon ``mass'' $m_\gamma^2 \approx e^3 B$ \cite{3, 33}. One would
expect that the photon mass should screen Coulomb potential and
shift energies of the atomic levels found in tree approximation.

Dirac equation spectrum in a constant homogenious magnetic field
looks like \cite{4}:
\begin{equation}
\varepsilon_n^2 = m^2 + p_z^2 + (2n+1)eB + \sigma eB \;\; ,
\label{1}
\end{equation}
where $n = 0,1,2, ...$, $\sigma = \pm 1$ and the field is directed
along axis $z$.\footnote{This spectrum with the substitution
of $2n+1+\sigma$ by $2j, \,\,j=0,1,2,... $ was found by I.I.
Raby \cite{IR}.} In the magnetic fields we are interested in
$\varepsilon_n \ga m/e$, and electrons are ultrarelativistic. The
only exception is the lowest Landau level (LLL) which has $n=0$,
$\sigma = -1$. The energy of LLL electron equals its mass for $p_z
= 0$ and the consideration of the nonrelativistic electron motion
along $z$ axis is selfconsistent. LLL is interesting both
practically and theoretically. An analog of the critical electric
field $E_{cr} = m^2/e$ is the magnetic field $B_0 = m^2/e =
4.4 \cdot 10^{13}$ gauss. Two orders larger superstrong fields
$B\ga m^2/e^3$ can exist at special neutron stars named magnetars.
The temperature of an outer magnetar layer is not enough to
populate the excited Landau levels and one can observe the
transitions among the states to which LLL is splitted at the
electric field of the nucleus. Freezing of the ground state energy
in the superstrong magnetic fields discussed in the paper leads to
the upper bound on the spectra of photons radiated from
magnetars.\footnote{I am grateful to S.I. Blinnikov for the
discussions of magnetar physics.} To study the stability of the
huge magnetic fields \cite{5} one should also know the energy of
the ground state as a function of a field.

So we are studying the energies of the states to which LLL splits
in the presence of an atomic nucleus.

Since the electron at LLL moves along $z$ axis we will study in
section 2 QED at $D=2$: the behaviour of electrons in
two-dimensional space-time. The coupling constant $g$ has
dimension of mass, so Coulomb potential as a function of $|z|$
depends on two dimensionfull parameters: $g$  and electron mass
$m$. We will obtain the approximate analytical formula for Coulomb
potential in $d= D-1 = 1$ which takes into account the photon
polarization operator. We will see that for large $g$ (or small
$m$) $g \gg m$ Coulomb potential is screened.

In section 3 we will consider the physical case, $D =4$ QED. The
analog of the coupling constant squared $g^2$ in the real world is
the product $e^3 B$. The polarization operator in the magnetic fields
$B\gg m^2/e$ at $k_\parallel ^2(\equiv k_z^2) \ll eB$ practically
coincides with the one obtained in section 2 \cite{77}.
Nevertheless the screening at large distances $|z| \gg 1/m$ does
not occur: at $|z| \gg 1/m$ we get a purely Coulomb potential
$\Phi(z) = e/|z|$. The screening occurs at small distances, and
its influence on the ground state energy is determined in section
4. The results similar to those presented in sections 3 and 4 were
obtained in \cite{55} with the help of the numerical calculations.

\section{$D=2$ QED: screening}

Coulomb potential in the coordinate space could be obtained by the
Fourier transformation of the 00-component of the photon
propagator in momentum representation at momentum $k_\mu=(0,
k_\parallel)$. We designate the space-like component of momentum
by $k_\parallel$, which will be natural in the case $d=3$, see
below. The series of Feynman diagrams for the photon proparator is
shown in Fig. 1 which corresponds to the following equations:
$$
\Phi(\bar k) \equiv A_0(\bar k) = \frac{4\pi g}{\bar k^2} \; ;
\;\; {\bf\Phi} \equiv {\bf A}_0 = D_{00} + D_{00}\Pi_{00}D_{00} +
...
$$

\newpage

\begin{center}
\bigskip
\includegraphics[width=.8\textwidth]{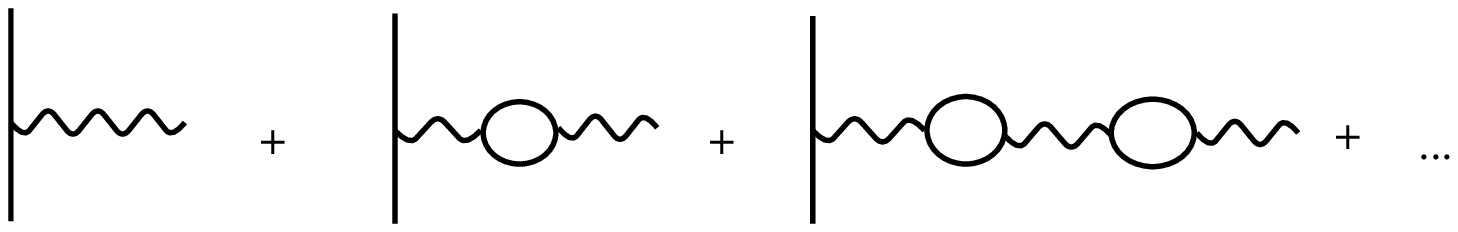}

Fig 1. {\em Modification of the Coulomb potential due to the
dressing of the photon propagator.}

\end{center}

Summing the series we get:
\begin{equation} {\bf\Phi}(k) = -\frac{4\pi g}{k^2
+ \Pi(k^2)} \; , \;\; \Pi_{\mu\nu}
\equiv\left(g_{\mu\nu}-\frac{k_\mu k_\nu}{k^2}\right)\Pi(k^2) \;\;
, \label{2}
\end{equation}
where $\Pi_{\mu\nu}$ is the photon polarization operator at one
loop. Instead of calculating the fermion loop we can take an
expression for $\Pi$ obtained in the dimensional regularization
method \cite{AB}, substitute $D=2$ in it and divide it by two,
because in two dimensions the traces of $\gamma$-matrices are
proportional to 2 instead of 4:
\begin{equation}
\Pi(k^2) = 4g^2\left[\frac{1}{\sqrt{t(1+t)}}\ln(\sqrt{1+t} +\sqrt
t) -1\right] \equiv -4g^2 P(t) \;\; , \label{3}
\end{equation}
$$
t\equiv -k^2/4m^2 \;\; \mbox{\rm --}
$$
-- a well-known result; for example see \cite{77}.
\footnote{It was demonstrated in \cite{77} that in strong magnetic
fields photon polarization operator is dominated by 
 the electron from LLL
and is essentially given by the $D=2$ expression.}

Taking $k = (0, k_\parallel)$, $k^2 = -k_\parallel ^2$ for the
Coulomb potential in the coordinate representation we get:
\begin{equation}
{\bf\Phi}(z) = 4\pi g \int\limits^\infty_{-\infty} \frac{e^{i
k_\parallel z} dk_\parallel/2\pi}{k_\parallel^2 + 4g^2
P(k_\parallel^2 /4m^2)} \;\; , \label{4}
\end{equation}
and the potential energy for the charges $+g$ and $-g$ is finally:
\begin{equation}
V(z) = -g{\bf\Phi}(z) \;\; . \label{5}
\end{equation}

The calculation  of ${\bf\Phi}(z)$ would be simplified if we were
interested in the correction to the potential $\sim g^2$.
Expanding denominator of (\ref{4}) and taking into account the
first two terms we would deform the integration contour in the
plane of complex $k_\parallel$ in such a way, that the integration
result will be given by the residue at $k_\parallel = 0$ and
integration of discontinuity of $P(k_\parallel^2)$ which equals
the imaginary part of it. This is analogous to what is done in the
textbook \cite{6} when the Uehling--Serber correction to Coulomb
potential in  $d=3$ is calculated. However to obtain the photon
mass we should take into account the whole infinite series -- mass
is not generated in a finite order of the perturbation theory.
Discontinuity of the integrand of (\ref{4}) is not equal to ${\rm
Im} P$ and the simplification of the integration does not occur.

Asymptotics of  $P(t)$ are:
\begin{equation}
P(t) = \left\{
\begin{array}{lcl}
\frac{2}{3} t & , & t\ll 1 \\
1 & , & t\gg 1 \;\; .
\end{array}
\right. \label{6}
\end{equation}

Let us take as an interpolating formula for $P(t)$ the following
expression:
\begin{equation}
\overline{P}(t) = \frac{2t}{3+2t} \;\; . \label{7}
\end{equation}

We have checked that the accuracy of this approximation is not worse
than 10\% for the whole interval of $t$ variation, $0 < t <
\infty$. Substituting(\ref{7}) in (\ref{4}) we easily perform the
integration:
\begin{eqnarray}
{\bf\Phi}(z) & = & 4\pi g\int\limits^{\infty}_{-\infty} \frac{e^{i
k_\parallel z} d k_\parallel/2\pi}{k_\parallel^2 +
4g^2(k_\parallel^2/2m^2)/(3+k_\parallel^2/2m^2)} = \nonumber
\\
& = & \frac{4\pi g}{1+ 2g^2/3m^2}
\int\limits_{-\infty}^{\infty}\left[\frac{1}{k_\parallel^2} +
\frac{2g^2/3m^2}{k_\parallel^2 + 6m^2 + 4g^2}\right]
e^{ik_\parallel z} \frac{dk_\parallel}{2\pi} = \\
&=& \frac{4\pi g}{1+ 2g^2/3m^2}\left[-\frac{1}{2}|z| +
\frac{g^2/3m^2}{\sqrt{6m^2 + 4g^2}} {\rm exp}(-\sqrt{6m^2
+4g^2}|z|)\right] \;\; . \nonumber \label{8}
\end{eqnarray}

In the case of heavy fermions ($m\gg g$)  the potential is given
by the tree level expression; the corrections are suppressed as
$g^2/m^2$:
\begin{equation}
{\bf\Phi}(z)\left|
\begin{array}{l}
~~  \\
m \gg g
\end{array}
\right. = -2\pi g|z|\left(1+O\left(\frac{g^2}{m^2}\right)\right)
\;\; . \label{9}
\end{equation}

In case of light fermions ($m \ll g$) the second term which
describes Yukawa potential in $d = 1$ dominates at the distances
$|z| < (1/g)\ln(g/m)$. At larger distances the first term
dominates; a coupling constant is suppressed by the factor
$3m^2/2g^2$ with respect to the tree level expression:
\begin{equation}
{\bf\Phi}(z)\left|
\begin{array}{l}
~~  \\
m \ll g
\end{array}
\right. = \left\{
\begin{array}{lcl}
\pi e^{-2g|z|} & , & z \ll \frac{1}{g} \ln\left(\frac{g}{m}\right) \\
-2\pi g\left(\frac{3m^2}{2g^2}\right)|z| & , & z \gg \frac{1}{g}
\ln\left(\frac{g}{m}\right) \;\; .
\end{array}
\right. \label{10}
\end{equation}

The dependence of the potential energy of the two opposite charges
(\ref{5}) on the distance between them is shown in Figure
2.\footnote{I am grateful to A.V. Smilga who noted privately that
in the case of light fermions in $D=2$ QED a massive pole in a
photon propagator emerges.}

\bigskip

\begin{center}

\bigskip
\includegraphics[width=.4\textwidth]{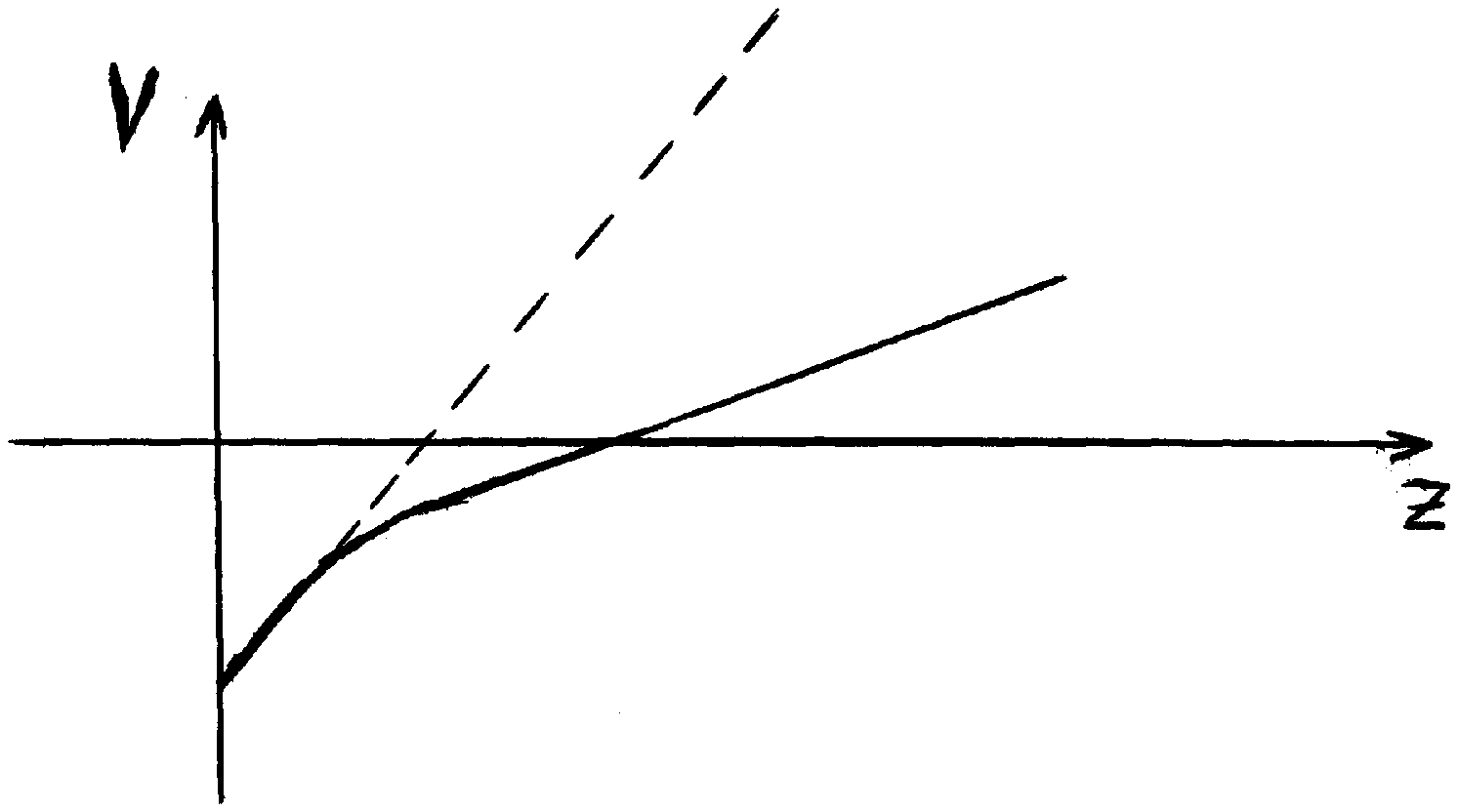}

Fig. 2. {\em The potential energy of two opposite charges in �
$D=2$ QED in the case $g \gg m$. The dashed line shows $V(z)$ for
$g\ll m$.}

\end{center}

In case $m=0$ a linear term disappears and the potential is
determined by the photon with the mass $m_\gamma = 2g$ exchange
(the Schwinger model:  $D = 2$ QED with massless fermions 
\cite{Sch}). For
massive fermions at the distances larger than $\ln(g/m)/g$ we
obtain a linear potential with the coupling constant diminished by the
factor $3m^2/2g^2$ .

\section{$D = 4$ QED in a strong magnetic field: screening at $|z| <
1/m$}

In order to derive the potential of the pointlike charge in the
realistic case of  $D=4$ QED with $d=3$ space-like dimensions in
the external magnetic field we should know the expression for a
polarization operator in this field. There are many papers where
the polarization operator in the external homogeneous field was
calculated, see \cite{7} -- \cite{9}. The expression for the
polarization operator radically simplifies in the magnetic field
which is so strong, that the Landau levels spacing is considerably
larger than the electron mass, $B \gg B_0 = m^2/e$ and at the
longitudinal (parallel to the magnetic field) photon momentum
$k_\parallel^2 \ll eB$, see Eqs.(1.19), (1.22) and (5.2) in
\cite{9}. With the help of these formulas we get:
\begin{eqnarray}
{\bf\Phi}(k)  & = & - \frac{4\pi e}{k^2 +\chi_2(k^2)} = \nonumber
\\
& = & \frac{4\pi e}{(k_\parallel^2 +
k_\bot^2)\left(1+\frac{\alpha}{3\pi}\ln\left(\frac{2eB}{m^2}\right)\right)
+ \frac{2 e^3 B}{\pi} {\rm exp}\left(-\frac{k_\bot^2}{2eB}\right)
P\left(\frac{k_\parallel^2}{4m^2}\right)} \; , \label{11}
\end{eqnarray}
where $k = (0, k_x, k_y, k_z)$, $k_\bot^2 = k_x^2 + k_y^2$, $k_z =
k_\parallel$ and the magnetic field is directed parallel to the
$z$ axis. This formula is very similar to the analogous formulas
from the previous section, see Eqs. (\ref{2}), (\ref{3}). The
difference is in an extra small term $\sim \alpha \ln(eB/m^2)$,
which we will not take into account in the future consideration,
and in the factor $2e^3 B/\pi {\rm
exp}\left(-\frac{k_\bot^2}{2eB}\right)$ which substitutes $4g^2$.
The dependence of function $P$ on  $k_\parallel^2$ is the same as
that in the case of $D=2$ QED. There is also an extra term
$k_\bot^2$ in the denominator and to obtain the potential in the
coordinate representation we should integrate over $k_\bot$ as
well.

With the help of interpolating formula $P(t)$ from section 2 we
obtain:
\begin{equation}
{\bf\Phi}(z) = 4\pi e \int\frac{e^{ik_\parallel z} d k_\parallel
d^2 k_\bot/(2\pi)^3}{k_\parallel^2 + k_\bot^2 + \frac{2 e^3B}{\pi}
{\rm exp}(-k_\bot^2/(2eB))(k_\parallel^2/2m^2)/(3+k_\parallel
^2/2m^2)} \;\; , \label{12}
\end{equation}
where the integration is performed in the cylindrical coordinates
and we are looking for the potential along axis $z$, since it is
the potential which bounds an electron in the atom.

We manage to find the asymptotic behaviour of ${\bf\Phi}(z)$ for
$z$ much larger and much smaller than Compton wave length of the
electron. For large distances $|z| \gg \frac{1}{m}$ in the
integral (\ref{12}) the region $|k_\parallel| \ll m$ is important
and for the magnetic field $B\gg B_0$ we get $k_\parallel^2 \ll
eB$ and the expression for $P$ we are using is correct. For small
$|k_\parallel|$ we get:
\begin{equation}
{\bf\Phi}(z)\left|
\begin{array}{l}
~~  \\
|z|\gg \frac{1}{m}
\end{array}
\right. = 4\pi e \int \frac{e^{ik_\parallel z} dk_\parallel d^2
k_\bot/(2\pi)^3}{k_\parallel^2 \left[1+\frac{e^3 B}{3\pi m^2} {\rm
exp}\left(-\frac{k_\bot^2}{2eB}\right)\right] +k_\bot^2} \; .
\label{13}
\end{equation}

It is convenient to integrate over $k_\parallel$ closing the
integration contour in an upper (lower) semiplane of the complex
$k_\parallel$ and taking $k_\parallel = ik_\bot/\sqrt{1+\frac{e^3
B}{3\pi m^2} {\rm exp} (-\frac{k_\bot^2}{2eB})}$. In this way we
obtain:
\begin{equation}
{\bf\Phi}(z)\left|
\begin{array}{l}
~~  \\
|z|\gg \frac{1}{m}
\end{array}
\right. = e \int\limits_0^\infty \frac{{\rm exp}\left[-k_\bot
|z|/\sqrt{1+\frac{e^3 B}{3\pi m^2} {\rm exp}\left(
-\frac{k_\bot^2}{2eB} \right)}\; \right]}{\sqrt{1+\frac{e^3
B}{3\pi m^2} {\rm exp}\left(-\frac{k_\bot^2}{2eB}\right)}}d k_\bot
\; . \label{14}
\end{equation}

The integral over $k_\bot$ converges at $k_\bot \la \sqrt{e^3 B}$,
that is why the residue was situated at $k_\parallel \ll m$, where
the approximate formula for $P$ we used is valid. At the mentioned
values of $k_\bot$ the exponent inside the square root can be
substituted by one and finally we obtain:
\begin{equation}
{\bf\Phi}(z) \left|
\begin{array}{l}
~~  \\
|z|\gg \frac{1}{m}
\end{array}
\right. = \frac{e}{|z|} \; , \;\; V(z)\left|
\begin{array}{l}
~~  \\
z\gg \frac{1}{m}
\end{array}
\right. = -\frac{e^2}{|z|} \;\; {\mbox {\rm --}} \label{15}
\end{equation}
-- the usual Coulomb potential. Strong screening which we obtain
in $d=1$ at the distances $|z| \gg 1/m$ in a realistic case $d=3$
does not occur.

To find a potential at short distances $|z|\ll \frac{1}{m}$ let us
substitute $m=0$ in (\ref{12}):
\begin{eqnarray}
{\bf\Phi}(z) \left|
\begin{array}{l}
~~  \\
|z|\ll \frac{1}{m}
\end{array}
\right. & = & 4\pi e \int\frac{e^{ik_\parallel z} dk_\parallel d^2
k_\bot/(2\pi)^3}{k_\parallel^2 + k_\bot^2 + \frac{2e^3 B}{\pi}
{\rm exp}\left(-\frac{k_\bot^2}{2eB}\right)} = \nonumber \\
& = & e\int\limits_0^\infty\frac{{\rm exp}\left(-\sqrt{k_\bot^2 +
\frac{2e^3 B}{\pi} {\rm
exp}\left(-\frac{k_\bot^2}{eB}\right)}|z|\right)}{\sqrt{k_\bot^2
+\frac{2 e^3 B}{\pi} {\rm exp}\left(-\frac{k_\bot^2}{eB}\right)}}
k_\bot dk_\bot \;\; . \label{16}
\end{eqnarray}
Calculating the potential at $|z|\gg 1/\sqrt{eB}$ we observe that
the integral over $k_\bot$ is determined by the integrand at
$k_\bot \ll \sqrt{eB}$. So we took residue at $k_\parallel \approx
k_\bot \ll \sqrt{eB}$ and the approximate expression for $P$ was
used in the domain where it is valid. Performing integration over
$k_\bot$ we get:
\begin{eqnarray}
{\bf\Phi}(z)  \left|
\begin{array}{l}
~~  \\
\frac{1}{m}\gg z \gg \frac{1}{\sqrt{eB}}
\end{array}
\right. & = & e\int\limits_0^\infty\frac{{\rm
exp}\left(-\sqrt{k_\bot^2 + \frac{2 e^3
B}{\pi}}|z|\right)}{\sqrt{k_\bot^2 +\frac{2 e^3 B}{\pi}}}
k_\bot dk_\bot = \nonumber \\
& = & \frac{e}{|z|} {\rm exp}\left(-\sqrt{\frac{2
e^3 B}{\pi}}|z|\right) \;\; , \nonumber \\
V(z) & = & -\frac{e^2}{|z|} {\rm exp} \left(-\sqrt{\frac{2 e^3
B}{\pi}}|z|\right) \;\; . \label{17}
\end{eqnarray}
At the distances which are smaller than Compton wave length we
obtain screening of the potential which corresponds to the photon
mass $m_\gamma^2 = 2 e^3 B/\pi$. Coulomb potential is screened for
the superstrong magnetic fields $B > m^2/e^3$.

\section{The energy of the ground state of the hydrogen atom
in the superstrong magnetic fields $B > m_e^2/e^3$}

According to papers \cite{1,2} in the magnetic fields $B
> m^2 e^3$ the ground state energy of the hydrogen atoms equals
$E_0 = -(me^4/2)\ln^2(B/m^2 e^3)$ and at $B = m^2/e^3$ it equals
$E_{\rm cr} = -(me^4/2)\ln^2(1/e^6)$. For larger magnetic fields
the screening of the Coulomb potential at the distances $|z|\la
\frac{1}{m}$ occurs. Let us demonstrate that the screening leads
to the freezing of the energy -- it does not diminish with the
growth of the magnetic field.

To find the ground state energy we use the following equation
\cite{2}:
\begin{equation}
E_0 = -2m\left(\;\int\limits_{a_H}^{a_B} U(z) dz\right)^2 \;\; .
\label{18}
\end{equation}
We split the integral into two parts: from $1/m$ to $a_B$, where
the screening is absent (large $z$),
\begin{equation}
I_1 = -\int\limits_{1/m}^{a_B} \frac{e^2}{z} dz = -e^2
\ln\left(1/e^2 \right) \label{19}
\end{equation}
and from the Larmour radius $a_H =1/\sqrt{eB}$ to $1/m$, where the
screening occurs (small $z$):
\begin{equation}
I_2 = -\int\limits_{1/\sqrt{eB}}^{1/m}\frac{e^2}{z} {\rm
exp}(-\sqrt{e^3 B} z) dz = -e^2 \ln(1/e) \;\; . \label{20}
\end{equation}
Finally we get:
\begin{equation}
E_0 = -m e^4/2 \ln^2(1/e^6) = -me^4/2 \times 220 \label{21}
\end{equation}
and the energy level freezing occurs. The numerical estimates of
Shabad and Usov give $73.8 \times 4 \approx 295$ instead of 220,
see Eq. (14) in Phys. Rev. Lett. paper \cite{55}.

When $B$ increases further Larmour radius approaches the size of a
proton. This happens at $1/\sqrt{eB} \approx 1/m_\rho$, $m_\rho =
770$ MeV, $B \approx 10^{20}$ gauss. Taking into account the
proton formfactor we get that for larger fields $I_2$ does not
contribute to the energy, factor 220 in (\ref{21}) should be
substituted by 100: the ground level goes up.

Without screening $I = -e^2\ln(a_B/a_H)$, $E_0 =
-(me^4/2)\ln^2(B/m^2 e^3)$ as it was stated in the beginning of
this section.

\section{Conclusions}

The photon polarization operator leads to modifications of the
atomic energy levels. The famous example is its contribution to
the Lamb shift, the difference of the energies  of $2s_{1/2}$ and
$2p_{1/2}$ levels of hydrogen. They are numerically small loop
corrections to the values of energies determined by the tree level
potential. The role of the photon polarization diagram in the
superstrong magnetic fields $B > m^2/e^3 = 6.2 \cdot 10^{15}$
gauss is qualitatively different. It determines the behaviour of
the ground state energy: the formula obtained at tree level becomes
invalid and the growth of the coupling energy with $B$ terminates
at $B\approx m^2/e^3$. Screening of Coulomb potential
should be more important for the energies of even excited states
which are more sensitive to the shape of the potential at small
distances \cite{155}. Degeneracy of even and odd excited states 
in the limit $B\Longrightarrow\infty$ is not lifted by
the screening. We study the analogy of the electric
potential in $d=1$ QED with massive electrons and in $d=3$ QED in
strong magnetic fields $B > B_0 = m^2/e$ which originates from the
coincidence of the polarization operators in these cases. A simple
analytical expression which equals the polarization operator with
10\% accuracy enables us to obtain an approximate formula for the
electric potential of the point charge in $d=1$ QED with massive
fermions and asymptotics of the potential in $d=3$ QED. In $d=1$
QED for a coupling constant $g$ larger than a fermion mass $m$ a
tree level formula is modified at $|z|> 1/g$. In $d=3$ QED a tree
level formula is modified at the distances $1/m> |z| > 1/\sqrt{e^3
B}$ while at large distances $|z|> 1/m$ Coulomb law is valid.

Analogous results for $D=4$ were obtained in \cite{55}. 

The other aspect of the Coulomb potential in the strong magnetic
field is investigated  in paper \cite{10}: it is supposed that
fermions obtained their mass due to a magnetic field (dynamical
fermion mass).

I am grateful to S.I. Blinnikov, V.A. Novikov, L.B. Okun, V.S.
Popov, and A.V. Smilga for useful discussions and to A.I.Rez,
who brought to my attention Phys. Rev. publication \cite{55}. 

This work was supported by the grants RFBR 08-02-00494,
Nsh-4172.2010.2 and by the contract of the RF Ministry of Science
and Education No. 02.740.11.5158.

\end{document}